\documentclass[aps,prb,showpacs,amsmath,amssymb]{revtex4}

\usepackage{graphicx,amsmath}

\date{\today}

\newcommand{\ba}{\begin{eqnarray}}
\newcommand{\ea}{\end{eqnarray}}
\newcommand{\be}{\begin{equation}}
\newcommand{\ee}{\end{equation}}
\newcommand{\eps}{\epsilon}
\newcommand{\mrm}[1]{\mathrm{#1}}
\newcommand{\mcal}[1]{{\mathcal{#1}}}
\newcommand{\drm}{\mrm{d}}
\newcommand{\ddt}{\frac{\partial}{\partial\,t}}

\newcommand{\ev}[1]{\langle {#1} \rangle}

\begin{document}

\title {Many-Body Dynamics and Exciton Formation Studied
by Time-Resolved Photoluminescence}
\author{W.~Hoyer}
\author{C.~Ell}
\author{M.~Kira}
\author{S.~W.~Koch}
\affiliation{Department of Physics and Materials Sciences Center,
Philipps-University Marburg, Renthof 5, 35032 Marburg, Germany}
\author{S.~Chatterjee}
\author{S.~Mosor}
\author{G.~Khitrova}
\author{H.~M.~Gibbs}
\affiliation{Optical Sciences Center, The University of Arizona,
Tucson, Arizona 85721-0094}
\author{H.~Stolz}
\affiliation{Department of Physics, University of Rostock,
Universit\"{a}tsplatz 3, D-18051, Rostock, Germany}
\date{\today}

\begin{abstract}
The dynamics of exciton and electron-hole plasma populations 
is studied via time-resolved photoluminescence after nonresonant
excitation. By comparing the peak emission at the exciton resonance
with the emission of the continuum, it is possible to experimentally
identify regimes where the emission originates predominantly from
exciton and/or plasma populations. The results are supported
by a microscopic theory which allows one to extract the fraction
of bright excitons as a function of time. 
\end{abstract}

\maketitle

\section{Introduction}
%
Excitons, i.e.~bound states between a conduction band electron and a 
valence band hole, are often regarded as the semiconductor analogue of
bound hydrogen atoms. In contrast to the hydrogen atom, however, the
exciton binding energy lies in the range of 10\,meV due to the small 
masses and the large background dielectric constant\cite{Haug:94}.
Furthermore, typical excitation conditions in semiconductors lead to
phase-space filling and screening such that for elevated carrier densities
the Coulomb interaction becomes screened and the binding even weaker.
Thus, only at low lattice temperatures and up to moderate densities
can bound exciton populations be expected to be the
dominating species in the interacting electron-hole system.
In general, the many-body state is a complicated mixture
containing Coulomb-correlated electrons and holes as well as
bound and ionized exciton correlations and populations.

Even more importantly, under typical experimental situations the real 
hydrogen atom in its ground state is stable against decay
while excitons first have to be generated by optical excitation 
and subsequent many-body interactions, and then they can recombine
radiatively or nonradiatively. It is known that the radiative lifetime 
in low-dimensional systems is usually much shorter than in 
the respective bulk materials\cite{feldmann,Andreani:91}.
Consequently, after nonresonant excitation where the laser is tuned 
to energies corresponding to the continuum of electron-hole transitions, 
it is interesting to find out if and how many excitons
form and which states they occupy.
The formation process involves Coulomb scattering and interaction
with lattice vibrations (phonons). Under favorable conditions, 
electron-hole pairs can give their excess energy to the phonon bath
or to the remaining carrier system and form bound excitons\cite{hoyer:03}.
Since the photon momentum is small compared to typical exciton
momenta, only excitons with extremely small momenta, i.e.\  
inside the light cone, can radiatively recombine. We refer to them 
as {\em bright excitons}. The remaining {\em dark excitons} first have 
to scatter down into the light cone before they can decay radiatively.
Hence, one would generally assume that under nonresonant
excitation conditions with subsequent exciton formation
the vast majority of these bound pairs would always exist in
the different dark states.

The finite radiative lifetime of bright excitons is on the order of 13\,ps 
at low temperatures in GaAs based single quantum-well systems. Monitoring
the time-dependent photoluminescence (PL) collected at the 1s-exciton
resonance therefore yields information about the dynamics of these
exciton populations. The build-up of the PL after nonresonant excitation
has been interpreted as build-up of an incoherent exciton
population \cite{kusano,blom,eccleston,damen,kumar,gulia}, and the
subsequent fall off was taken as evidence for this population
decay\cite{feldmann,deveaud}. However, a recently developed microscopic
luminescence theory for a Coulomb correlated plasma of electron-hole
pairs predicts that luminescence peaks at the 1s-exciton resonance
even without the presence of incoherent exciton populations\cite{kira98}.
In fact, the cooling dynamics of an unbound electron-hole plasma
after interband excitation leads to a luminescence spectrum whose
peak grows and sharpens at the 1s-exciton resonance as the carriers
relax to the bottom of their respective bands. Accordingly, the 
mere existence of PL at the spectral position of the 1s-resonance 
would not prove the existence of an incoherent
population of excitons, and previous interpretations using a
purely excitonic picture need to be reviewed carefully.

Since the energy differences between exciton states correspond to
frequencies in the terahertz regime, terahertz absorption
measurements provide a powerful alternative method
to study exciton formation\cite{kaindl,chari}. Theoretically,
it has been shown that the build-up of the induced absorption 
corresponding to the exciton 1s to 2p transition is sensitive
only to the correlated exciton populations and gives thus direct
evidence for exciton populations\cite{kira01,kira04}. Current results
from terahertz absorption measurements after nonresonant excitation
suggest exciton formation times of several hundreds of picoseconds\cite{kaindl}.

A recent publication based on photoluminescence measurements combined with
a microscopic theoretical model for emission from plasma and excitons\cite{chatter} 
shows that only a small fraction of the carrier density is transformed into
optically active excitons under best formation conditions. However, even
this small bright 1s-exciton population can lead to massive changes
of the emission spectra. Clearly, since the PL is not 
sensitive to dark excitons, these measurements do not allow to
unambiguously determine the total number of excitons in the system. 
Nevertheless, it seems to be established by now that 
under suitable density and excitation conditions an
exciton population is present at a liquid-helium lattice
temperature ($\approx 4$\,K) and that this population may dominate the 
1s-emission properties. The precise formation dynamics and in particular
the interpretation of time resolved PL measurements, however, is still
controversial. For example, a recent experiment \cite{hayes} shows that in 
nonresonantly excited time resolved PL measurements the 1s-resonance is 
developed on a sub-ps timescale at 100\,K, much faster than any expected
exciton formation time. 
We interpret this experiment as direct support for the predicted
emission at the 1s-resonance originating from populations of unbound, Coulomb 
correlated electron-hole plasma states.

In the present work, we complement our published experimental results\cite{chatter} 
by a detailed study of time-resolved PL combined with nonlinear 
absorption measurements to determine the carrier density. After an
outline of the experimental and theoretical methods in Sec.~\ref{methods}
we discuss our results in Sec.~\ref{experiment}. We show that 
from the investigation of the dynamics of the experimental PL spectra 
alone, one can clearly distinguish two regimes with and without 
major exciton contributions. This confirms our previous result
that above 30\,K the emission at the 1s-resonance is dominantly due to
unbound pair-state emission, while for temperatures lower than 30\,K an
increasing optically active exciton population contributes\cite{chatter}.
We support our experimental results by a microscopic theory 
leading to the luminescence analogue of the famous Elliott
formula~\cite{elliott}. It includes not only the exciton
resonances but also a source term which contains both a plasma
contribution and an incoherent exciton population. This theory is
used to extract a time resolved number of bright excitons.

\section{Methods}
\label{methods}
%
\subsection{Experiment}
\label{expmethods}
%
The principal sample (DBR42) consists of twenty
In$_{0.06}$Ga$_{0.94}$As quantum wells. Each is 8\,nm thick and
grown in between 130\,nm GaAs barriers. The indium concentration
and well thickness were chosen to place the 1s heavy-hole (hh)
exciton resonance at 1.471\,eV at 4\,K lattice temperature. The
linewidth of the exciton transition is 0.96\,meV full width at half maximum (FWHM);
the exciton binding energy is 8\,meV. This sample permits the detection
of about 13\,meV of emission above the 1s hh exciton resonance,
undistorted by substrate absorption and impurities. At the same
time, the 1s-transition energy is high enough to lie within
the operating range of laser and detector. We excited nonresonantly
13.2\,meV above the 1s-resonance, into the heavy-hole continuum
but below the light-hole resonance. Both sides of the sample were
anti-reflection coated to reduce Fabry-P\'{e}rot interference
fringes. The experimental results were checked on several other
samples.

Photoluminescence and nonlinear absorption measurements were
performed under identical conditions, i.e.\ sample temperature
and excitation density. The linear and nonlinear
absorption of the sample under investigation was measured with a ps
pump-probe setup. Both pulses were generated by a solid-state
pumped, actively mode-locked Titanium:Sapphire laser with a fixed
repetition rate of 80\,MHz and a maximum time-integrated output
power of 2\,W. The peak wavelength of the spectrally broad 100\,fs 
probe pulse, tunable using a birefringent filter, was typically
centered at the heavy-hole continuum edge.

The long PL lifetime of several nanoseconds
required a reduction of the laser repetition rate. For the
experiments presented here, an electro-optic modulator reduced it
to 2\,MHz to eliminate carrier accumulation effects in the sample.
To measure the absorption corresponding to PL signals that take
11\,ns to decay, it was necessary to expand and collimate the probe 
beam and pass it twice through a $\approx$\,1\,m delay line. For selective
excitation, a spectrally narrower tunable 3\,ps pump pulse was
generated in a grating pulse shaper. The spectral width of the
pump pulse was 2\,meV FWHM. The sample was held in a cold finger
cryostat. The pump and probe pulses were focussed onto the sample
using 10\,cm and 45\,cm focal length lenses, yielding 60\,$\mu$m
and 20\,$\mu$m spot sizes, respectively. The transmission and
PL were spectrally resolved using identical imaging
grating monochromators with a spectral resolution of 0.8\,meV. The
transmitted probe light was detected using a liquid nitrogen cooled
charged-coupled device (CCD) camera, and the PL with
a Hamamatsu streak camera with a slow sweep unit in single photon
counting mode. This mode of operation reduces the time resolution 
to 90\,ps. To even better utilize the dynamic range of the streak camera, 
a film neutral density filter was placed at the exit plane of the
monochromator reducing the fluence by a factor of 40 for all
energies $<$ 1.474\,eV. This was extremely helpful because a partial
goal of this study was to detect several meV of continuum emission
which is intrinsically four to five orders of magnitude lower than
the emission peak at the 1s-resonance. All data shown here have
been corrected for this attenuation and are presented undistorted.
Photoluminescence spectra are integrated from t-50\,ps to t+50\,ps
around each time step.

To estimate the carrier density $n_{\mrm{eh}}$ during the decay at 
time $t$ after the pump pulse, $n_{\mrm{eh}}(t)$ is assumed to be linearly 
proportional to the change in absorption $\Delta\alpha$L at time $t$. From
the absorption at the pump energy, the measured pump power and the pump spot
size at the sample, the initial carrier density was estimated and the
corresponding nonlinear absorption was measured at 10\,ps before any reduction 
occurred. This calibration curve linking $\Delta\alpha$L and $n_{\mrm{eh}}$
was used to determine the carrier density at any time delay. This calibration 
might not be completely unambiguous, since the nonlinear absorption
spectrum is influenced differently by carriers bound to excitons, 
$n_{\mrm{X}}$, as compared to quasi-free carriers, $n_{\mrm{eh}}-n_{\mrm{X}}$. 
However, since in the following experiments in general 
$n_{\mrm{X}} \ll n_{\mrm{eh}}$, we neglect the subtle differences and believe
that the nonlinear absorption still provides an accurate measure of the total 
carrier density.

\subsection{Theory}
\label{theomethods}
%
In order to interpret the experimental observations quantitatively,
we apply our microscopic quantum theory. In the current approach we utilize
findings from previous publications \cite{chatter,hoyer:03} which
allow us to compute the PL spectrum under quasi steady-state
conditions and to clearly distinguish between bound and unbound
pair state contributions. The theory starts from the fundamental
Hamiltonian including the quantized carriers coupled via Coulomb
interaction, and interacting with phonons and a quantized light field.
This theory, evaluated at the level of a Hartree-Fock approximation,
first predicted PL at the exciton energy without exciton
populations \cite{kira98}. Meanwhile we have extended the analysis to
include also electron-hole correlations and 
bound excitons \cite{kira01,hoyer:03}.

The fundamental quantity for the computation of a PL
spectrum is the rate of emitted photons,
%
\begin{eqnarray}
I_{\rm PL}(\omega_q)
&=&
\ddt \Delta\ev{B^{\dagger}_{q} B_{q}}
=
\frac{2}{\hbar} \mrm{Re}\left[ \sum_{k} \mcal{F}_q^\star
\Delta\ev{B^{\dagger}_q a^{\dagger}_{v,k} a_{c,k}} \right],
\label{eq:BdagB}
\end{eqnarray}
%
where $B^{\dagger}_{q} B_{q}$ is the photon number operator for photons with wavenumber $q$,
$a^{\dagger}_{\lambda,k}$ creates an electron in band $\lambda = c,v$ in quantum state $k$, and
$\mcal{F}_q$ denotes the coupling matrix element to mode $q$. We have restricted the analysis to
emission perpendicular to the quantum well such that the wave vector $q$ is a scalar quantity
and equivalent to the frequency $\omega_q = c_0\,q$. In Eq.~(\ref{eq:BdagB}), the $\Delta\ev{\dots}$
denotes the correlated (i.e.\ incoherent) part and is given by the full expectation value minus
its factorized (i.e.\ classical) contribution \cite{Wyld:63,Hoyer:04b}.

The photon number and thus the PL are coupled to photon-assisted polarizations of the
form $\Delta\ev{B^{\dagger} a^{\dagger}_v a_c}$. Therefore, we set up the equation of motion for
this quantity,
%
\begin{eqnarray}
\ddt \Delta\ev{B^{\dagger}_q a^{\dagger}_{v,k} a_{c,k}}
& = &
(\eps_k -\hbar \omega_q - i \gamma^{\mrm{D}}_{k}(\omega_q) ) 
\Delta\ev{B^{\dagger}_q a^{\dagger}_{v,k} a_{c,k}}
-
(1 - f^e_k - f^h_k ) \sum_{k'} V_{k-k'} \Delta\ev{B^{\dagger}_q a^{\dagger}_{v,k'} a_{c,k'}}
\nonumber\\
&&\quad
+ i \sum_{k'} \gamma^{\mrm{OD}}_{k,k'}(\omega_q) \Delta\ev{B^{\dagger}_q a^{\dagger}_{v,k'} a_{c,k'}}
+ i \mcal{F}_q (f^e_k f^h_k + \Delta\ev{a^{\dagger}_{c,k'} a^{\dagger}_{v,k} a_{c,k} a_{v,k'}}),
\label{eq:Bv+c}
\end{eqnarray}
%
where we have neglected a stimulated contribution which can describe radiative coupling between
multiple quantum wells and is not important under the experimental conditions. Here,
$\eps_k $ is the total kinetic energy of electron and hole in state $k$, and $f^{e(h)}_k$
describes the microscopic carrier distribution of electrons (holes). Furthermore, we have
included Coulomb scattering via $ \gamma^{\mrm{OD}}_{k,k'}(\omega_q)$ and
$\gamma^{\mrm{D}}_{k}(\omega_q)$. These scattering matrices are obtained
from solving and inserting the higher order correlations
$\Delta\ev{B^{\dagger}a^{\dagger}_v a^{\dagger}_\lambda a_\lambda a_c} $ in second-Born
approximation. The last term of Eq.~(\ref{eq:Bv+c}) provides
the source for spontaneous emission. It contains the Hartree-Fock contribution proportional
to the product of electron and hole distributions \cite{kira98} and also a correlated source.
This correlated source is a true two-particle expectation value and changes depending on
whether electrons and holes form a correlated plasma or whether additional bound excitons
are present. This term will be discussed in detail below.

In order to calculate the PL spectrum, we use an adiabatic treatment for the
photon-assisted polarizations and introduce a generalized Wannier basis which
diagonalizes the homogeneous part of Eq.~(\ref{eq:Bv+c}). Due to the phase-space filling
factor $(1 - f^e_k - f^h_k )$ and the presence of the second-Born scattering matrices,
the respective matrix is not Hermitian and we have to distinguish between right- and
left-handed eigenfunctions $\phi^{l/r}_{\nu,q}$. Furthermore, we obtain complex
eigenvalues of the form $E_{\nu}(\omega_q) - i \gamma_{\nu}(\omega_q)$. Both real
and imaginary part of these eigenvalues depend on the photon energy through the
energy denominator in the second-Born scattering rates. Since the imaginary part
of the eigenvalue depends on the excitonic state, this basis also correctly
describes the fact that energetically higher lying exciton states show nonlinear behavior
in the form of a larger linewidth for much lower densities than the lowest bound state.
Using this exciton basis, we obtain the steady-state PL as
%
\begin{eqnarray}
I_{\rm PL}(\omega_q) = \frac{|d_{\rm cv}^2|\omega_q}{\varepsilon_{\rm bg}}
{\rm Im}\Bigl[
     \sum_{\nu} \frac{\phi^{r}_{\nu,q}(r=0)}{E_{\nu}(\omega_q) - \hbar\omega_q - i \gamma_{\nu}(\omega_q)}
\sum_{k,k'} [\phi_{\nu,q}^{l}(k)]^\star\, \ev{a^{\dagger}_{c,k'}a_{v,k'} a^{\dagger}_{v,k}  a_{c,k} }
\Bigr].
\label{eq:lumi1}
\end{eqnarray}
%
The prefactor in Eq.~(\ref{eq:lumi1}) is determined by the square of the dipole
matrix element $|d_{\rm cv}|^2$ and the background dielectric
constant $\varepsilon_{\rm bg}$ \cite{param}.
Equation~(\ref{eq:lumi1}) is reminiscent of the famous Elliott
formula for bandgap absorption \cite{elliott}; it contains a sum
over exciton states, and the resonances of the denominator show
that the PL peaks at the same excitonic energies as the
absorption. In fact, a solution of the nonlinear susceptibility
similar to the method above leads to
%
\begin{eqnarray}
\chi(\omega_q) = \frac{|d_{\rm cv}^2|}{\varepsilon_{\rm bg}}
{\rm Im}\Bigl[
     \sum_{\nu} \frac{\phi^{r}_{\nu,q}(r=0)}{E_{\nu}(\omega_q) - \hbar\omega_q - i \gamma_{\nu}(\omega_q)}
\sum_{k} [\phi_{\nu,q}^{l}(k)]^\star\,(1 - f^e_k - f^h_k)
\Bigr] ,
\label{eq:abs}
\end{eqnarray}
%
with the same excitonic basis $\phi_{\nu,q}$, energies $E_{\nu}(\omega_q)$,
and damping constants $\gamma_\nu(\omega_q)$. Due to the frequency dependence
of the $\gamma_{\nu}(\omega_q)$, Eqs.~(\ref{eq:lumi1}) and~(\ref{eq:abs}) can 
describe non-Lorentzian lineshapes. In the limit of low densities, the
$\gamma_{\nu}$ approach zero and the phase-space filling factor becomes
approximately 1, and we recover the low-density Elliott formula.
In contrast to the absorption, the strength of the PL according to
Eq.~(\ref{eq:lumi1}) is not only determined by the exciton wavefunctions but
also by the source term $\sum_{k,k'} [\phi_{\nu,q}^{l}(k)]^\star\,
\ev {a^{\dagger}_{c,k'} a_{c,k} a_{v,k'} a^{\dagger}_{v,k} }$.
This source contains a factorized contribution
%
\be
\sum_{k} [\phi_{\nu,q}^{l}(k)]^\star\, f^e_k f^h_k
=
\sum_{\nu'}[\phi^{r}_{\nu',q}(r=0)]^\star\, \ev{X^{\dagger}_{\nu',q} X_{\nu,q}}_{\mrm{HF}}
\label{eq:HFsource}
\ee
%
with
%
\be
\ev{X^{\dagger}_{\nu',q} X_{\nu,q}}_{\mrm{HF}}
=
\sum_k \phi^{l}_{\nu',q}(k)\,[\phi^{l}_{\nu,q}(k)]^{\star}  f^e_k f^h_k,
\label{eq:defHFsource}
\ee
%
which is always present as soon as electrons and holes are excited. This
is the source term due to which 1s-luminescence without exciton
populations has been
predicted first \cite{kira98}. Additionally, the correlated part of the
source can describe a correlated plasma as well as additional incoherent
bound exciton correlations $N_{\mrm{1s}}$ which may or may not be in the system.
From our previous results \cite{hoyer:03} we know that even without the
formation of any bound excitons, the two-particle correlations do not
vanish and describe a correlated plasma. Without the inclusion of the
second-Born scattering discussed above, it can be shown that
the presence of the correlations mainly cancels all the off-diagonal
contributions $\nu\not=\nu'$ in Eq.~(\ref{eq:HFsource}). In the
present paper, however, we numerically solve the full sum of
Eq.~(\ref{eq:HFsource}) and treat the correlated part explicitly.

Thus, the last missing input for the evaluation of Eq.~(\ref{eq:lumi1}) is
the two-particle correlations. The full equation of motion without Coulomb
scattering has been given in a previous publication \cite{hoyer:03}. For the
present theory-experiment comparison, we also include second-Born scattering
which is provided by an approximative solution of six-point quantities.
Similar to the photon-assisted quantities, the equation of motion for the
exciton correlations contains a homogeneous part and an external source
depending on the carrier distributions $f^e$ and $f^h$. We therefore proceed
similarly as with the photon-assisted correlations in that we solve the homogeneous
equation by transforming into an adapted exciton basis $\tilde\phi^{l/r}_\nu$.
This basis set is different from the one used for the photon-assisted
polarizations because for the pure particle correlations there is no dependence
on the optical frequency $\omega_q$ such that the second-Born scattering rates
and consequently the complex eigenvalues $\tilde{E}_{\nu} - i \tilde\gamma_{\nu}$
are different from the ones discussed above. Using this new basis we can express the
correlated source as
%
\be
\sum_{k,k'}[\phi_{\nu,q}^{l}(k)]^\star\,
\Delta\ev{a^{\dagger}_{c,k'} a_{v,k'} a^{\dagger}_{v,k} a_{c,k} }
= \sum_{\nu',\nu''} [\tilde\phi^r_{\nu'}(r=0)]^{\star}
\left(\sum_{k} [\phi_{\nu,q}^{l}(k)]^\star\,\tilde\phi^r_{\nu''}(k) \right)
\Delta\ev{X^{\dagger}_{\nu'} X_{\nu''} }
\label{eq:PLsource}
\ee
%
where $X_{\nu}$ is the exciton annihilation operator. The
term $\Delta\ev{X^{\dagger}_{\nu'} X_{\nu''} }$ can easily be obtained from
the adiabatic solution of the Heisenberg equation of motion. In the absence of
phonon scattering or other mechanisms leading to exciton formation it is given by
%
\be
\Delta\ev{X^{\dagger}_{\nu'} X_{\nu''} }_{\mrm{pl}}
= -\frac{S_{\nu',\nu''}}
{\tilde{E}_{\nu''} - \tilde{E}_{\nu'} -i (\tilde\gamma_{\nu''} + \tilde\gamma_{\nu'})}
\ee
%
with the single particle source $S_{\nu',\nu''}$ given by
%
\be
S_{\nu',\nu''}
= \sum_{k,k'} \tilde\phi_{\nu'}(k') [\tilde\phi_{\nu''}(k'')]^{\star} V_{k'-k''}
\left[
(1 - f^e_{k'} - f^h_{k'}) f^e_{k''} f^h_{k''} - (1 - f^e_{k''} - f^h_{k''}) f^e_{k'} f^h_{k'}
\right].
\ee
%
The proper inclusion of Coulomb scattering and the distinction between the different
basis sets for photon-assisted polarizations and correlations, respectively, is vital
for a quantitative theory-experiment comparison over the experimentally relevant
density regime. Everything mentioned so far is regarded by us as PL resulting
purely from unbound pair states
since in the end all terms depend solely on single-particle electron and hole
distribution functions. For the comparison with the experiment, we take the
experimentally estimated carrier density and temperature as input for these
single-particle distributions and calculate the emission. In addition,
e.g.\ in the presence of phonon scattering, extra bound 1s-excitons may form
as discussed in Ref.~\onlinecite{hoyer:03}. In the present paper, we do not include
the formation dynamics explicitly; instead, we add a purely exciton contribution
$N_{1s,q=0}$
%
\be
\Delta\ev{X^{\dagger}_{\nu'} X_{\nu''}}
=
\Delta\ev{X^{\dagger}_{\nu'} X_{\nu''} }_{\mrm{pl}}
+ \delta_{\nu',\mrm{1s}} \delta_{\nu'',\mrm{1s}} N_{1s,q=0}
\ee
%
of optically active 1s-exciton correlations in Eq.~(\ref{eq:PLsource}).
The exciton contribution is thus treated as input to the theory
and varied until agreement with the experimental spectra is reached.
Thereby, we can distinguish between bound and unbound pair-state populations
and extract the number of optically active ($q \approx 0$) 1s-excitons.

\section{Results}
\label{experiment}
%
\subsection{Correlated plasma regime}
\label{lowtemp}
\begin{figure}
\resizebox{0.5\textwidth}{!}{\includegraphics{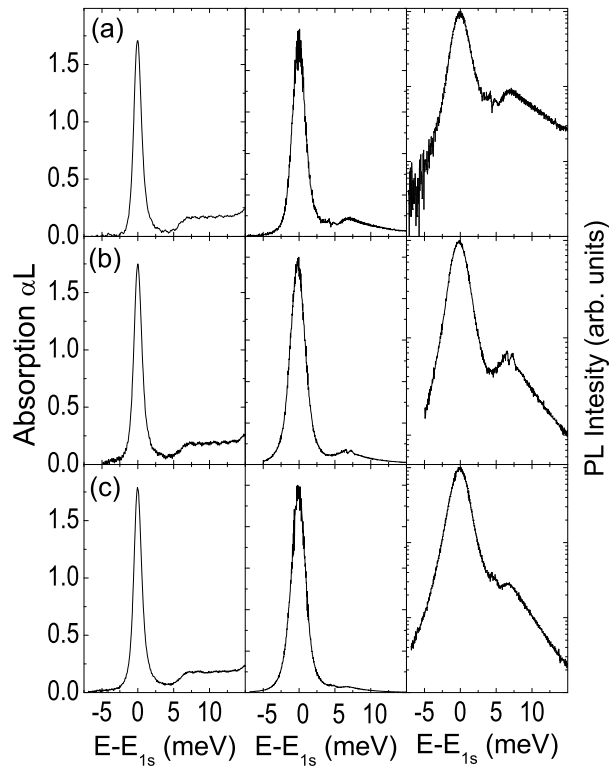}}
\caption{Nonlinear absorption spectra (left) and PL spectra on a
linear scale (middle) and logarithmic scale (right) for a time
delay of 1 ns for various lattice temperatures (a) 70\,K, (b) 50\,K, 
and (c) 30\,K. The initial carrier density is $n_{\mrm{eh}} = 3.3\times
10^{9}$\,cm$^{-2}$.}
\label{alplspectra}
\end{figure}

%
Nonlinear absorption measurements (left) and corresponding
PL spectra (middle and right) are shown in Fig.~\ref{alplspectra}
at a time delay of 1 ns after excitation.
%
%
The initial carrier 
density is $3.3 \times 10^{9}$\,cm$^{-2}$, and the nominal
lattice temperatures are (a) 70\,K, (b) 50\,K and (c) 30\,K. 
The absorption spectra (left) exhibit a strong absorption at the 
1s-exciton resonance followed by a 2s peak and the onset of the 
electron-hole continuum.
To allow for easier comparison of the spectral shape,
the energy scale is chosen relative to the 1s-exciton absorption
peak at 1.4650\,eV (70\,K), 1.4681\,eV (50\,K), and 1.4697\,eV (30\,K).
The line broadens only slightly in this temperature range \cite{angela}.

The corresponding PL spectra are shown on a linear scale
(middle) and on a semi-logarithmic scale (right). The emission is
strongly peaked at the 1s-exciton resonance and clearly dominates
the linear luminescence spectra. The 2s emission is seen as well. 
A small discontinuity on the high energy slope between 3\,meV and 
7\,meV above the 1s peak is visible in the PL spectra; it is
due to the imperfect correction of the attenuator used for lower
energies. While the main features of the emission stay the same,
namely that the spectrum is dominated by a peak at the 1s-resonance 
for all shown lattice temperatures, the most obvious
difference is the decreasing slope of the continuum emission with
increasing temperatures.

Coulomb scattering causes the electrons and holes to thermalize to
Fermi-Dirac distributions with a carrier temperature $T$, often
much higher than the lattice temperature $T_{L}$, on a
sub-picosecond time scale \cite{Knox86,Knox88}. Therefore, 
within the achieved time resolution, thermalization can 
be regarded as instantaneous. To investigate the carrier
temperature further we assume that the electrons and holes in the
continuum are in thermal equilibrium. In thermodynamic equilibrium,
the Kubo-Martin-Schwinger (KMS) relation \cite{kubo57,martin59}
%
\be
I_{\mrm{PL}}^{\mrm{eq}}(\hbar \omega) \propto g(\hbar \omega -
\mu)\,\alpha(\hbar \omega)
\label{eq:KMS}
\ee
%
is sometimes used to establish the relation between the absorption
coefficient $\alpha$ and the PL. Here, $\mu$ is the joint chemical
potential of the electrons and holes, and $g(E) = (\exp(\frac{E}{k_BT}) - 1)^{-1}$ 
is the Bose distribution function. For sufficiently low densities,
the Bose function can be approximated by a Boltzmann distribution
function $g(E) \approx A\,\exp(-\frac{E}{k_BT})$. This approximation is 
justified for the density and
temperature range investigated here. Thus the emission decreases
exponentially above the band edge, and the carrier temperature can
be extracted from the slope of the continuum emission. Due to
experimental noise, a small spectrally and temporally flat
background may be present in the data; it is subtracted in all
data shown. The data are therefore fit using a least mean square
fit for the variables, amplitude A and carrier temperature T, with
the corresponding standard deviations $\sigma _A$ and $\sigma _T$.
%
\begin{figure}
\resizebox{0.5\textwidth}{!}{\includegraphics{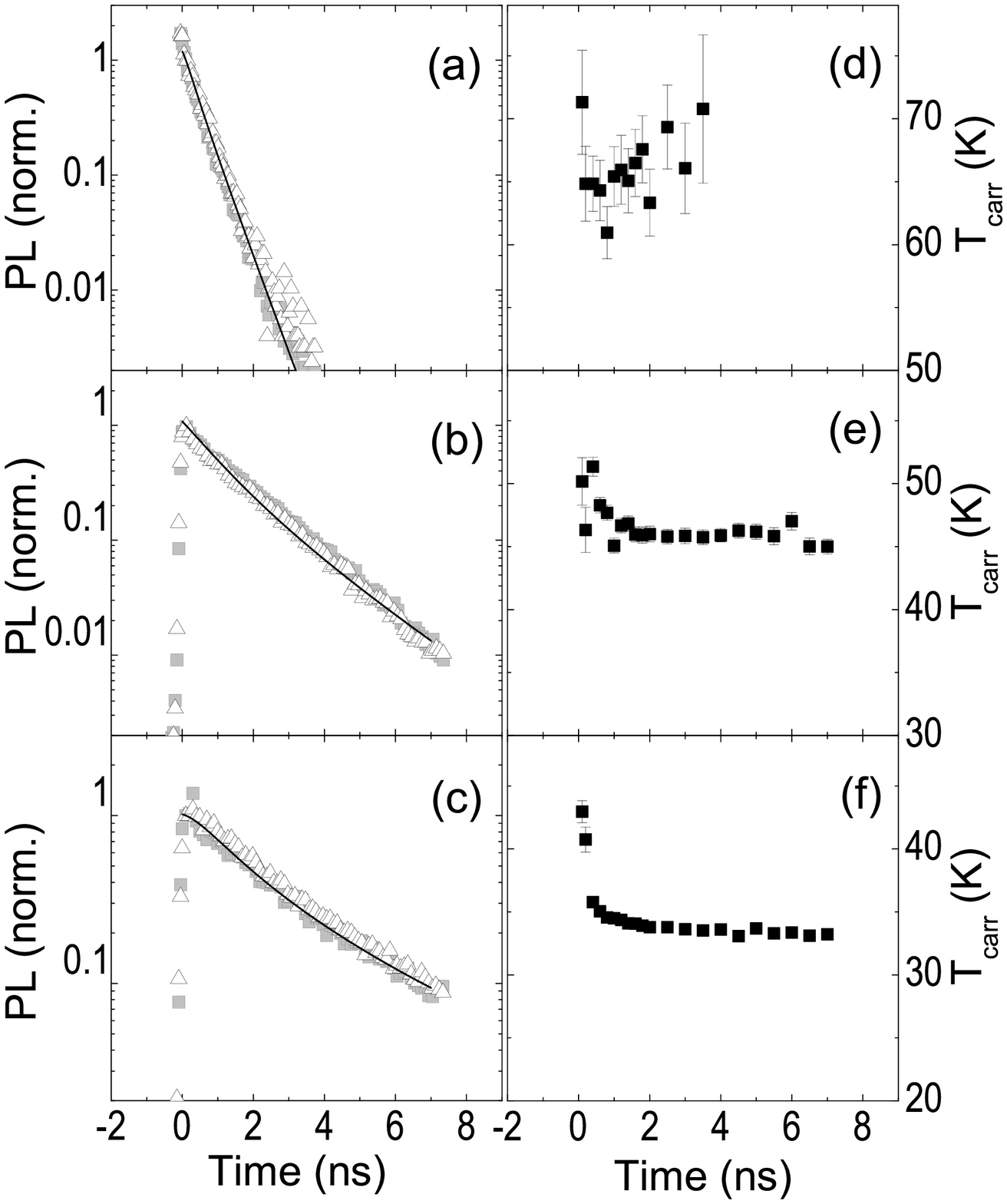}}
\caption{Temporal evolution of the PL of the 1s-exciton resonance
(open triangles) and of the continuum edge (full squares) for an
initial carrier density of $n_{\mrm{eh}} = 3.3\times 10^{9}$\,cm$^{-2}$ at
lattice temperatures of (a) 70\,K, (b) 50\,K, and (c) 30\,K. Fits
using the simplified plasma model are shown as solid lines. The
corresponding temporal evolution of the carrier temperature is
shown on the right side at lattice temperatures of 70\,K (d),
50\,K (e) and 30\,K (f).} 
\label{TRspectra}
\end{figure}
%
Figures~\ref{TRspectra}(d), (e), and (f) show the temperatures extracted 
from the PL data under the conditions of Fig.~\ref{alplspectra} that displays 
the 1\,ns spectra. The carrier temperatures approach values close to the lattice
temperature in all three cases. Times $\leq$ 0.1 ns are not accessible because 
of limited time resolution of the experiment; also the concept of a carrier 
temperature is ambiguous because of highly nonequilibrium Coulomb scattering
processes at early times.

The rise and subsequent decay of the PL were detected
time-resolved. The time evolution of the 1s-emission can be
studied in several ways. It is possible to spectrally integrate
between the FWHM points or all energies below the midpoint between
1s and 2s. Alternatively, the peak of the emission spectrum can be
traced. Here the spectrum is integrated 50\,$\mu$eV around the peak
of the linear absorption spectrum. Use of any of the other
methods of analysis does not change the results significantly. The
continuum emission is spectrally integrated 
from the band edge to 100\,$\mu$eV above it to achieve an acceptable
signal to noise ratio. A representative set of normalized emission
spectra can be seen in Fig.~\ref{TRspectra} (a)--(c) for nominal
lattice temperatures of 70\,K (a), 50\,K (b), and 30\,K (c), shown
on a semi-logarithmic scale. Compared within each plot are the
emission from the 1s-exciton resonance (open triangles) and the
emission from the continuum edge (full squares).

Clearly, the emission from the 1s-exciton resonance and the continuum 
emission have the same temporal dependence for lattice temperatures
of 70\,K and 50\,K, strongly indicating that the emission source is
the same in both cases. The peak of the continuum and 1s PL
is reached shortly after the end of the excitation.
For a lattice temperature of 30\,K, 1s-emission and continuum
emission start to deviate slightly: The emission from the 1s-exciton 
resonance continues to increase slightly after the end of
the excitation pulse. This deviation becomes more and more
pronounced for lower lattice temperatures, as discussed in the
next section.

To investigate the time evolution of the PL further for situations where
exciton populations can be omitted, we apply a simple rate equation model. 
Under such conditions, the Elliott formula Eq.~(\ref{eq:lumi1}) predicts a 
plasma source term $S_{\mrm{pl}} \propto n_{\mrm{eh}}^2(t)$ in the low density 
limit. Therefore, the continuum emission reduces to the well known 
$I_{PL} \propto B\,n_{\mrm{eh}}^2(t)$, where $B$ is the electron-hole bimolecular
recombination rate. Accordingly, the temporal evolution of the carrier density is
governed by the bimolecular recombination rate:  $dn_{\mrm{eh}}/dt = - B\, n_{\mrm{eh}}^2$. 
However, the PL in the experimentally investigated
time range cannot be fitted by only using the radiative decay proportional to $B$;
a nonradiative recombination $\propto -A\,n_{\mrm{eh}}$ must be included,
which describes the exponential like decay at longer times. As the
temperature is increased, the nonradiative decay becomes faster,
which can be at least partially explained by carriers escaping
from the quantum well; especially the holes escape easily due
to their shallow quantum well. Subsequently, carriers may decay
nonradiatively and emit PL elsewhere.
                                                                                
The corresponding fits using the simplified model without exciton populations
are included in Fig.~\ref{TRspectra} (a)--(c) as solid lines. As can
be seen in the figure, the simplified model describes
well the experimental data. The radiative recombination
coefficient B agrees within a factor of 2 to 3 with the expression
given in Ref.~\onlinecite{blood}. Hence, the coinciding temporal
evolution of the 1s and the continuum emission together with the
well fitting model unambiguously point to the unbound
electron-hole plasma as the sole source of emission from both the
1s-exciton resonance and the electron-hole continuum for lattice
temperatures $T>$ 30\,K.

\subsection{Excitonic regime}
\label{verylowtemp}
\begin{figure}
\resizebox{0.7\textwidth}{!}{\includegraphics{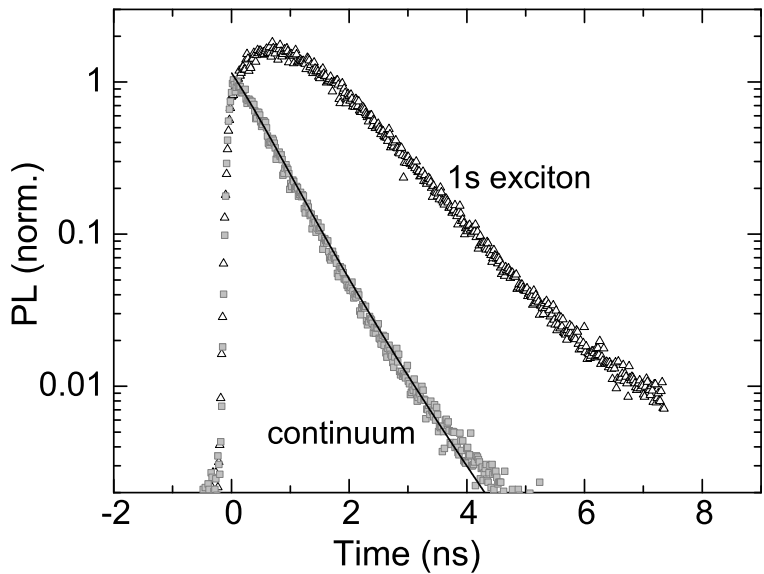}}
\caption{Temporal evolution of the PL of the 1s-exciton resonance
(open triangles) and of the continuum edge (full squares) for an
initial carrier density of $n_{\mrm{eh}} = 3.3\times 10^{9}$\,cm$^{-2}$ at
a lattice temperature of 4\,K. The fit using the simplified plasma
model is shown as solid line.}
\label{4KTRspectra}
\end{figure}

%
Lowering of the lattice temperature further leads to a changing
picture. At 4\,K lattice temperature, the time evolution of the
emission from the 1s-exciton resonance differs significantly from
that of the continuum: The maximum of the peak 1s-emission is
delayed compared to that of the continuum emission; see
Fig.~\ref{4KTRspectra}. 
%
%
For lower excitation densities, the maximum
is reached about 0.8\,ns after excitation. With increasing carrier
densities, the maximum occurs earlier. For the highest
investigated density the luminescence maximum is reached after
0.4\,ns. Thus, the 1s PL develops its own dynamics and
cannot be solely described by emission from unbound carriers; luminescence
from an incoherent exciton population also contributes\cite{chatter}.

To further analyze and quantify the amount of bright excitons
contributing to the 1s PL, we introduce a single
parameter $\beta$ \cite{chatter,schnabel92}, defined as
%
\be
\beta = \frac{I_{\mrm{PL}}(\mrm{1s})}{I_{\mrm{PL}}^{\mrm{eq}}(\mrm{1s})} .
\ee
%
Here, $I_{\rm PL}(\mrm{1s})$ is the measured PL at the 
1s-resonance or the calculated one by applying our microscopic
theory. $I_{\rm PL}^{\mrm{eq}}(\mrm{1s})$ satisfies the KMS relation
Eq.~(\ref{eq:KMS}). Experimentally and theoretically, $I_{\mrm{PL}}^{\mrm{eq}}(\mrm{1s})$
is found by multiplying the measured and calculated nonlinear $\alpha$ by a
Boltzmann factor, using the temperature extracted from the
measured continuum emission, and normalizing it to agree with the
measured and calculated continuum PL. Thus, $\beta$ quantifies how
the 1s-emission of a given spectrum differs from that expected
from the measured absorption assuming validity of the KMS
relation, with the measured and KMS PL spectra forced to agree
in the continuum.

%
\begin{figure}
\resizebox{0.5\textwidth}{!}{\includegraphics{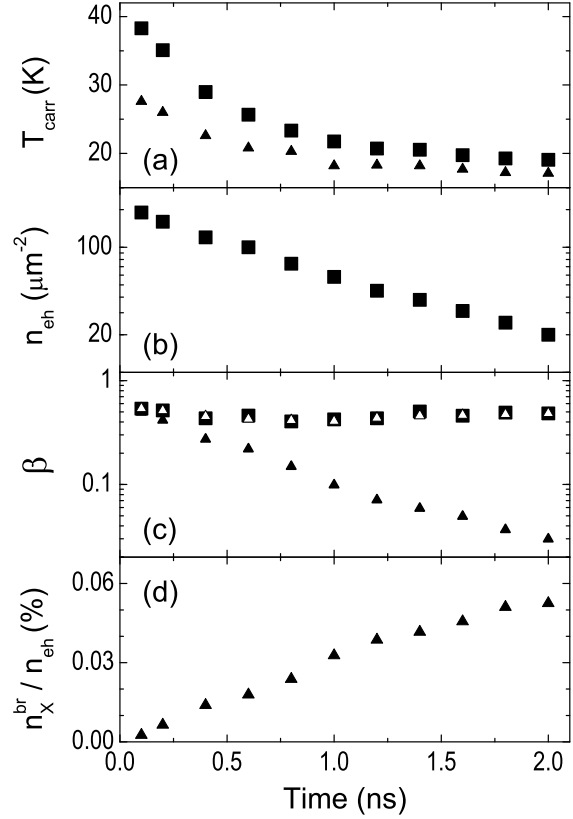}}
\caption{(a) Temperature dynamics as extracted from the experimental PL 
spectra (squares) and carrier temperatures used as input for the theory (triangles).
The lattice temperature is 4\,K. (b) Temporal evolution of the experimentally 
deduced carrier density. The initial carrier density is 
$n_{\mrm{eh}} = 1.9\times 10^{10}$\,cm$^{-2}$. 
(c) Parameter $\beta$ as function of time: experimentally (full
squares), calculated with (open triangles) and without (full
triangles) inclusion of an exciton population. (d) Extracted
fraction of bright excitons.}
\label{analysis}
\end{figure}
%
The theory-experiment comparison is shown in Fig.~\ref{analysis}
for an initial carrier density of $1.9 \times 10^{10}$\,cm$^{-2}$.
The temporal evolution of the experimentally deduced carrier
density is plotted in Fig.~\ref{analysis} (b).
Figure~\ref{analysis} (a) shows the temporal evolution of the
carrier temperature extracted from the continuum tail of the
corresponding measured luminescence spectra (full squares). For
all investigated carrier densities the electron-hole plasma cools
exponentially with typical decay times of 0.5 to 0.6\,ns . However, 
in contrast to the data series taken at higher lattice temperatures, 
the carrier temperature never reaches the lattice
value of 4\,K~\cite{schnabel92,leo88,yoon96,szczytko04}.

The input parameters to our theoretical analysis are the carrier
temperature and density. While the density can directly be taken 
from the experiment, the extracted temperature from the tail of the 
continuum PL is not necessarily identical to the electronic temperature which 
enters the theory. Our theory shows that as a consequence of Coulomb scattering
the extracted temperature from the continuum emission is always slightly 
higher than the carrier temperature put into the simulation.
Therefore, our analysis has to be done self-consistently and we
have run simulations with carrier temperatures adjusted in such
a way that the resulting continuum tail of the spectrum agrees with the experiment. 
The carrier temperatures put into the simulations 
are shown as full triangles in Fig.~\ref{analysis}~(a). 
We note that especially for early times 
with relatively high carrier density and corresponding stronger 
scattering the deviation between carrier temperature and extracted 
continuum slope is appreciable and stronger than for later times.

Figure~\ref{analysis}~(c) displays the time dependence of $\beta$.
Experimental values are plotted as full squares. The corresponding
theoretical values of the bare plasma emission are shown as full
triangles. They decrease as time goes by as a consequence of the
decreasing carrier density. This is in full agreement with our
previous findings. As discussed in Ref.~\onlinecite{chatter},
the calculation without exciton populations underestimates the
measured $\beta$ and an optically active $q = 0$ 1s-exciton
population has to be added, which strongly enhances the 1s PL.
Excitons are added until the theoretical and experimental $\beta$'s
are equal\cite{steady-state}; see open triangles in Fig.~\ref{analysis}~(c).

Since PL and thus $\beta$ are only sensitive to bright excitons, only
the number $N_{\mrm{1s},q=0}$ of 1s-excitons with vanishing center-of-mass
momentum can be directly obtained from our theory. In order to extract the 
bright exciton density $n_{\mrm{X}}^{\mrm{br}}$ from $N_{\mrm{1s},q=0}$, we approximate 
the integral
%
\be
n_{\mrm{X}}^{\mrm{br}} = \frac{1}{4\pi^2}\int_{|q_{\|}|<q_{\mrm{max}}} \!\!\!N_{\mrm{1s},q} \,\drm^2q
\approx N_{\mrm{1s},q=0} \frac{q_{\mrm{max}}^2}{4\pi},
\ee
%
where the maximum value for the parallel component of the wave vector,
$q_{\mrm{max}}$, depends on the experimental set up, the spot size, etc.
In the present paper, in accordance with Ref.~\onlinecite{chatter}, we count
all excitons with $q_{\mrm{max}} < \frac{E_g}{\hbar c_0}$, where $E_g$ is the
gap energy and $c_0$ the vacuum velocity of light. Thus, all excitons emitting
photons which can leave the barrier are counted as bright excitons.
Alternatively, we could have used $q_{\mrm{max}} < \frac{E_g n}{\hbar c_0}$,
corresponding to all excitons which can emit photons into the substrate with refractive
index $n$. However, since all photons with $\frac{E_g}{\hbar c_0} < |q_{\|}| < \frac{E_g n}{\hbar c_0}$
experience total internal reflection at the substrate-air boundary, they
can easily be reabsorbed by the quantum well and do not necessarily contribute
to a decrease in exciton density.

The calculated fraction of bright excitons $n_{\mrm{X}}^{\mrm{br}}/n_{\mrm{eh}}$ necessary to
reproduce the experimental $\beta$ values is plotted in Fig.~\ref{analysis}~(d).
The temporal evolution clearly demonstrates that --- for our material system and 
excitation conditions --- excitons need several hundreds of ps to form and 
subsequently to relax down to the optically active $q \approx 0$ state, where $q$ is the exciton
center-of-mass momentum. This time span is in full agreement with
recent THz results and theoretical predictions \cite{kaindl,hoyer:03}.
We want to emphasize that the fraction of bright excitons is displayed. While this
fraction monotonuously increases over 2\,ns, the bright exciton density $n_{\mrm{X}}^{\mrm{br}}$
itself exhibits a maximum around $t=800$\,ps. As an additional caveat, we also
note that the value of $\beta$ is extremely sensitive to the temperature 
deduced from the
continuum PL. Therefore, the theoretically determined bright exciton numbers 
have been extracted carefully and are assumed to be correct, 
but due to temperature and other parameter uncertainties,
the exact values could possibly be off by at most a factor of two.

%
\begin{figure}
\resizebox{0.5\textwidth}{!}{\includegraphics{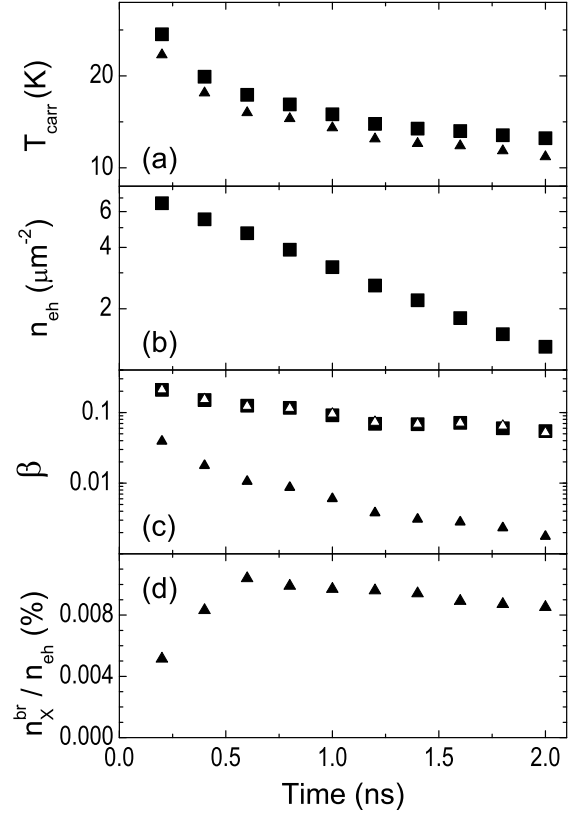}}
\caption{Same as Fig.~\ref{analysis} but for an initial carrier 
density of $n_{\mrm{eh}} = 7.9\times 10^{8}$\,cm$^{-2}$.}
\label{fig5}
\end{figure}
%
Figure~\ref{fig5} displays equivalent data as Fig.~\ref{analysis}, but now
for a much lower initial carrier density of $n_{\mrm{eh}} = 7.9\times 10^{8}$\,cm$^{-2}$.
While the qualitative behavior is very similar, a few differences must be
pointed out. First of all, the temperature extracted from the continuum PL
drops faster and approaches lower values than in the previous case of higher
density. Secondly, the density falls off more slowly. While in Fig.~\ref{analysis}
it drops by roughly a factor of 10 within 2\,ns, it now drops by less than
a factor of 7. 
As far as $\beta$ is concerned, the pure plasma value is smaller by an
order of magnitude compared to Fig.~\ref{analysis}. Nevertheless, fewer
excitons are needed to correctly fit the experimental data. Not only is
the absolute value of $N_{\mrm{1s},q=0}$ smaller, but even the fraction
of bright excitons shown in Fig.~\ref{fig5}~(d) is smaller and displays
a qualitatively different behavior than in Fig.~\ref{analysis}. At later
times, the excitons appear to decay quicker than the carrier density such
that the ratio $n_{\mrm{X}}^{\mrm{br}}/n_{\mrm{eh}}$ exhibits a maximum
after 600\,ps.

\section{Conclusions}
\label{conclusion}
%
Emission from the 1s-exciton resonance has been studied by means
of time-dependent PL measurements. A microscopic theory has been 
applied to quantify the amount of bright 1s-excitons. The regime of 
very low, liquid-helium lattice temperatures shows the presence of 
an incoherent population of excitons, which dominates
the 1s-emission dynamics in a wide range of carrier densities. A
time of 600 to 800 ps is found necessary for exciton formation and
relaxation to the lowest momentum states. With increasing
lattice temperature a smooth transition occurs at about 30\,K to a
regime where the 1s-emission is completely due to the population of 
unbound electron-hole plasma states.

\begin{acknowledgments}
%
The work is supported in Marburg by the Optodynamics Center of
the Philipps-Universit{\"a}t Marburg, in Rostock and Marburg
by the Deutsche Forschungsgemeinschaft through the Quantum Optics 
in Semiconductors Research Group,
and in Tucson by NSF (AMOP), AFOSR (DURINT), and COEDIP.
\end{acknowledgments}



\begin{thebibliography}{99}
\bibitem{param} Our calculations are performed for finite
quantum well widths and infinite barrier heights. We adjust the
effective well width such that our calculations reproduce the
experimentally measured exciton binding energy. $m_h/m_e = 3$ as
determined using $k \cdot p$ theory; the other material parameters
cancel out of the $\beta$ calculations.
\bibitem{steady-state} Since Coulomb scattering takes place on a fast sub picosecond
time scale, the application of our steady-state theory for extracting $\beta$ and
exciton numbers is justified for the times under display.
\end{thebibliography}
\end{document}